\documentclass[11pt,english]{article}
\pdfoutput=1
\usepackage{jheppub}
\usepackage[numbers]{natbib}
\usepackage{lscape}
\usepackage{amsmath}
\usepackage{slashed}
\usepackage{mathtools}
\usepackage{cleveref}
\usepackage{tcolorbox}

\usepackage{subcaption}
\usepackage{amssymb,tikz}
\usetikzlibrary{positioning}

\tikzset{sgplattice/.style={inner sep=1pt,norm/.style={red!50!blue},char/.style={blue!50!black},
  lin/.style={black!50}},cnj/.style={black!50,yshift=-2.5pt,left=-1pt of #1,scale=0.5,fill=white}}
\usetikzlibrary{shapes.geometric, arrows}

\usetikzlibrary{decorations.markings}
\tikzstyle{startstop} = [rectangle, rounded corners, 
minimum width=3cm, 
minimum height=1cm,
text centered, 
draw=black, 
fill=red!30]

\tikzstyle{io} = [trapezium, 
trapezium stretches=true, 
trapezium left angle=70, 
trapezium right angle=110, 
minimum width=3cm, 
minimum height=1cm, text centered, 
draw=black, fill=blue!30]

\tikzstyle{process} = [rectangle, 
minimum width=3cm, 
minimum height=1cm, 
text centered, 
text width=3cm, 
draw=black, 
fill=orange!30]

\tikzstyle{decision} = [diamond, 
minimum width=3cm, 
minimum height=1cm, 
text centered, 
draw=black, 
fill=green!30]
\tikzstyle{arrow} = [thick,->,>=stealth]

\begin{document}

\title{An effective description of the instability of coherent states of gravitons in  string theory}

\author[a]{Cesar Damian} 
\author[b]{and Oscar Loaiza-Brito}

\affiliation[a]{Departamento de Ingenier\'ia Mec\'anica, Universidad de Guanajuato, Carretera Salamanca-Valle de Santiago Km 3.5+1.8 Comunidad de Palo Blanco, Salamanca, Mexico.}
\affiliation[b]{Departamento de F\'isica, Universidad de Guanajuato, Loma del Bosque No. 103 Col. Lomas del Campestre C.P 37150 Leon, Guanajuato, Mexico.}


\date{\today}

\abstract{We study the dynamics of a coherent state of closed type II string gravitons within the framework of the Steepest Entropy Ascent Quantum Thermodynamics, an effective model where the quantum evolution is driven by a maximal increase of entropy. We find that by perturbing the pure coherent state of gravitons by the presence of other coherent fields in the string spectrum, there exists conditions upon which the system undergoes decoherence by reaching  thermodynamical equilibrium. Following the proposal by Dvali, et al. \cite{Dvali:2017eba}, this suggests the instability of the classical dS space. We identify the time scale it takes the system to reach  equilibrium consisting of a mixed state of fields in the string spectrum and compare it with the quantum-break time. Also we find that in such final state the quantum-break time seems to be larger than the classical break-time, in agreement with the Swampland conjectures about the dS solution in string theory.
}
\arxivnumber{}


\maketitle

%
%

\section{Introduction}


The current accelerated expansion of our Universe is nicely described in a classical approximation by a 4-dimensional de-Sitter  (dS) space-time. However, its construction from fundamental structures in a quantum gravity theory, such as string theory, has been challenging. One of the first proposals to construct a stable dS vacuum was given almost 20 years ago by Kachru, Kallosh, Linde and Trivedi, known as the KKLT solution \cite{Kachru:2003aw}, which viability is still in debate at present day. There have been as well more unorthodox constructions as the incorporation of non-geometric fluxes \cite{Damian:2013dq, Damian:2013dwa, Leontaris:2023obe, AbdusSalam:2024arh}, which nevertheless present important limitations \cite{Damian:2023ote, Prieto:2024shz}.\\

The difficulty to find a stable fully controlled dS vacuum in string theory has open up the possibility of other scenarios. One of the most intriguing is the sugestion about the non-existence of dS in string theory \cite{Danielsson:2018ztv} or in any framework of quantum gravity \cite{Garg:2018reu, Obied:2018sgi, Ooguri:2018wrx} through the (Refined) dS conjectures within the Swampland Program \cite{Palti:2019pca, vanBeest:2021lhn, Grana:2021zvf, Agmon:2022thq} in which the existence of a stable dS solution is compromised at Planckian energies and in consequence at the string landscape. These proposals suggest in turn, that dS space could be unstable or metastable. \\

In order to describe the possible instability of  dS space, it was proposed in \cite{Dvali:2017eba} (see also \cite{Dvali:2018jhn, Berezhiani:2021zst, Berezhiani:2024boz}) that the classical dS space-time can be visualized,  at the weak gravity limit, as a coherent state of gravitons over the Minkowski space-time. As the theory incorporates non-linearities, the coherent dynamics departs from the classical dynamics, defining a breakdown of this relation at some time, called the {\it classical} break time $t_{cl}$, which for the case of dS space goes as
\begin{equation}
    t_{cl}\sim \frac{1}{H},
\end{equation}
with $H$  the Hubble constant. However, there is also a quantum dynamics leading to instability. In the presence of other particles interacting with the gravitons, the system undergo decoherence indicating that dS space is no longer described by a classical mean field evolution. The time at which happens is called the {quantum} break time $t_Q$ and goes as
\begin{equation}
    t_Q\sim \frac{M^2_{Pl}}{H^3}.
\end{equation}
 Therefore, dS space, identified as a coherent state of gravitons scattering to other particles, is unstable. The relation between both times goes as $t_{cl}\sim \alpha t_q$, where $\alpha=H^2/M^2_{Pl}$. The decoherence of dS space within the framework of string theory  as well as the connection with the swampland Refined dS conjecture was addressed in \cite{Blumenhagen:2020doa}.\\

In the present paper we follow this idea, by constructing a coherent state of gravitons in type II superstring theory and study its quantum evolution by an effective model which modifies the Schr\"odinger equation by adding a dissipation term. The evolution is then driven by the increase of entropy among the quantum states conforming the system. Specifically,  the coherent states, constituying pure quantum states,  can be dissipated by the presence of other orthogonal states, as typical happens in the closed string by the presence of the extra fields coming from the NS-NS sector, $B$-field, dilaton and the RR fields, all of them conforming the correspondent coherent states.\\

The referred framework consider  
a non-linear dynamics called the Steepest Entropy Ascent Quantum Thermodynamics (SEAQT) and it was proposed in  \cite{Beretta:2009rmp}. In this formulation, the system evolves not according to unitary, reversible dynamics, but instead in a direction that maximizes the rate of entropy production, constrained by the principles of quantum mechanics.  
In this scenario, the state of the system follows a trajectory of steepest entropy ascent, driving it toward thermodynamic equilibrium. This proposal   is an effective model which allows to describe how the initial coherent state of gravitons in string theory undergoes decoherence.\\

Hence, following \cite{Dvali:2017eba} and identifying a coherent state of gravitons as the quantum structure of a classical dS space at weak gravity limit, it is possible to study the existence of conditions upon which the dS space could be unstable by undergoing decoherence based on the effective parameters establish by the SEAQT. In this framework, the quantum interactions and scattering of gravitons are encoded in a dynamics driven by the ascent of entropy which in turn allows to compute the time scale in which the system can reach equilibrium by undergoing decoherence which describes how the coherent state of gravitons evolves into a mixed state conformed by all NS-NS and RR massless fields in the closed string. \\

Let us then summarize our main results:
\begin{enumerate}
    \item We effectively describe the evolution of a quantum system conformed by a coherent state of gravitons, using the SEAQT framework.
    \item In this context, the evolution is driven by the increase of entropy. For that it is essential to have orthogonal coherent states interacting with the initial state of coherent gravitons. This scenario is given in the closed superstring theories, where the RR fields are orthogonal to the graviton by constructions, and , as we shown in Appendix A, the rest of the NS-NS fields, forming a coherent state,  are also orthogonal to the graviton coherent state.
    \item By studying the effective evolution of the system, we find that there exists a time scale at which the system can reach equilibrium by undergoing decoherence and ending in a mixed state conformed by fields in the NS-NS and RR sectors.
    \item By identifying the coherent state of gravitons as the dS space we then conclude that dS is unstable. The time in which the classical dS space is not longer described by a dynamics of a quantum coherent state is called quantum-break time and denoted $t_Q$. We find that $t_Q$ is essentially related to the the difference of mean energy values among different coherent states. Also, in the equation of motion describing the effective evolution of the system, it appears a parameter in time units which seems to be independent of the quantum process but that nevertheless determines how long it takes the system to evolve into decoherence. We therefore identify this parameter with the classical break-time $t_{cl}$.
    \item Closed to equilibrium when the system is a mixed state, one can see that $t_Q>t_{cl}$, a condition that seems to forbid dS space to undergo a quantum process, since the classical-break time is the time scale at which the system is destroyed by classical non-linearities which apart the theory from the weak coupling limit. This is in agreement with the Swampland conjectures.\\
\end{enumerate}

The outline of our work is as follows. In section 2 we implement the so called Steepest-Entropy-Ascent Quantum Thermodynamics (SEAQT) framework to a system of coherent state of gravitons within the context of type II string theory and review the most important characteristics of this proposal. Section 3 is devoted to apply this framework to a pure state constructed from closed string gravitons in the presence of other states. We show how the decoherence is natural  due to the presence of orthogonal states such as the B-field $B_2$ and the dilaton $\phi$ in the NS-NS sector and all the massless states in the RR one. We show that such state is unstable and compute the time scale in which equilibrium is restored. Finally we shall discuss about some implications on the stability of the dS space when seen as a coherent state of gravitons over the Minkowski space and  the classical and quantum break time\footnote{While finishing this manuscript, we became aware of \cite{Brahma:2024ycc}, which share some common topics.}.






\section{The SEAQT framework}
In this section we describe the generalities about the framework we shall use to study the stability of a coherent quantum state.  It is important to notice that this framework is an effective proposal to describe the dynamical evolution of a quantum system, by assuming that it is driven by an increment of entropy. The model was proposed in \cite{beretta2006nonlinear} and it has been successful in providing an effective description of quantum systems by incorporating thermodynamic irreversibilities directly into the system dynamics. For instance, the SEAQT framework has successfully applied to predict a flip-change process, essential during heat interactions with an environment, and pure dephasing \cite{Montanez:2022decoherence,Montanez:2020loss}, which is related in this framework solely to internal entropy generation. It also models atom-photon field interactions in cavity quantum electrodynamics (CQED), accurately predicting the decay of coherence in photon fields and agreeing with experimental observations \cite{Cano:2015steepest} as well as the relaxation of non-resonant excitation transfer
processes \cite{Morishita:2023relaxation}.\\

Let us review the basics of this framework. Consider a nonlinear extension of the Schrödinger (von Neumann) equation, which incorporates a dissipative term inducing an increase on entropy. This method proposes that the system follows the path in Hilbert space along which the entropy increases most rapidly, providing a quantum channel compatible with the second law of thermodynamics \cite{Witten:2020mini}. Specifically, it involves projecting the gradient of the entropy onto the orthogonal direction  to a constrained space or manifold, spanned by the usual quantum dynamics without dissipative terms. This manifold is defined by the conserved observables, which act as generators of motion. Typical examples of these generators include the identity operator, the Hamiltonian operator $\hat{H}$, and particle number operators \cite{Beretta:2009rmp,beretta2006nonlinear,beretta2014steepest,beretta2020fourth}. For a single, indivisible system, this means that the usual symplectic dynamics is modified by including a nonlinear dissipative term, resulting in an evolution equation of the form
\begin{equation}
\label{eq:eom}
    \frac{d\hat \rho}{dt} = -\frac{i}{\hbar} \left[ \hat H , \hat \rho \right] + \hat{\mathbb{D}} \left( \hat \rho , \hat H \right),
\end{equation}
where the Hermitian operator $\hat{\mathbb{D}}$ describes the presence of dissipation associated to the quantum state and it is given by
\begin{equation}
    \hat{\mathbb{D}} \left( \hat{\rho}, \hat{H} \right) = -\frac{1}{2 \tau_{D}} \left( \sqrt{\hat{\rho}} \, \hat{D} + \text{h.c.} \right),
\end{equation}
where
$\hat \rho$ is the density matrix operator, and $\tau_D$ is a free parameter. Consequently $\sqrt{\hat \rho}$  is the square-root state operator obtained from the spectral expansion of the density matrix $\hat \rho$ by substituting its eigenvalues with their positive square roots. Finally,
 $\hat D$ encodes the dissipative process that we proceed to describe. The quantum dynamics of non-dissipative (i.e. a system where the entropy  does not increase) can be thought of a dynamics on a manifold generated by the basis of operators $\sqrt{\hat{\rho}}$ and $\sqrt{\hat{\rho}}\hat H$. The idea is then to consider an orthogonal direction, along which the quantum system takes a path (which inherently implies the increase of entropy). Therefore, one can say that $\hat{D}$ must be of the form 
\begin{equation}
    \hat{D}= \sqrt{\hat \rho}( \hat B \ln \hat \rho )-\left[ \sqrt{\hat \rho}\hat B \ln \hat \rho\right]_{\perp \mathcal{L}\{ \sqrt{R}_i\}},
    \end{equation}
such that the dissipative process is encoded in the operator $\hat{B}$, i.e., if $\hat{B}$ is the identity operator, we have a maximal dissipation where $\hat{D}$ is totally aligned along the orthogonal direction given by $\sqrt{\hat{\rho}}\log\hat{\rho}$ which is chosen in order to correctly reproduce the von Neumann entropy, as we shall shortly see. \\

For a system evolving to a state of larger entropy, the most general form of the operator $\hat{B}$ is
$\hat B = I - \hat P_{\text{Ker}\rho}$, where $\hat{P}_{\text{Ker}\rho}$ projects onto those states which have zero eigenvalues with respect to $\hat{\rho}$. 
This orthogonality conditions ensures that both trace and energy are preserved along all the dynamics. In this way, the  operator $D$ (which contrary to $\hat{\mathbb{D}}$, is non-hermitian) can be constructed by the Gram-Schmidt process  through the non-recursive formula
\begin{equation}
    \hat D = \frac{\left|
    \begin{matrix}
    -\sqrt{\hat{\rho}} \hat S & \sqrt{\hat{\rho}} & \sqrt{\hat{\rho}} \hat{H} \\
    -\langle S \rangle  & 1  & \langle H \rangle \\
    -\langle H S \rangle & \langle H \rangle & \langle H^2 \rangle
    \end{matrix}\right|
    }{\Gamma \left( \{ \sqrt{\hat{\rho}}, \sqrt{\hat{\rho}}\hat{H} \} \right)}
\end{equation}
where $\langle H \rangle = \text{tr} \left( \hat \rho \hat H\right)$ and  $\Gamma \left( \{ \sqrt{\hat \rho}, \sqrt{\hat \rho} \hat H \} \right)$ is the Gram determinant. For correctly reproducing the von Neumann entropy, we define 
 $\hat S = - \hat B \ln \hat \rho$ and as usual,
$\langle S \rangle= \text{tr}\left( \hat \rho \hat S \right)$. Therefore, the Hermitian operator takes the form
\begin{equation} \label{eq:dissp}
\mathbb{D} = - \hat{\rho} \hat S + \frac{\beta}{2} \left\{ \hat{\rho}, \hat{H} \right\} + \left( \langle S \rangle - \beta \langle H\rangle \right) \hat \rho,
\end{equation}
where $\{ , \}$ is the anti-commutator and where we use the shorthand notation $\langle \cdot \rangle = \langle \Phi | \cdot | \Phi \rangle $ for a general quantum state $|\Phi\rangle$. Notice that $\hat{\mathbb{D}}$ can be written as the difference between the average free energy and the operator free energy as
\begin{equation}
    \mathbb{D} = \beta \left( \hat f - \hat \rho \langle f \rangle \right)
\end{equation}
with $\langle f \rangle = \text{tr} \hat f$, 
\begin{equation} \label{eq:freeop}
    \hat f = \frac{1}{2} \left\{ \hat \rho, \hat H \right\} - \frac{ \hat \rho \hat S}{\beta}
\end{equation}
and the non-equilibrium inversed temperature $\beta$  given by
\begin{equation}
    \beta = \frac{\langle \hat H \hat S \rangle - \langle \hat H \rangle \langle \hat S \rangle}{\langle \hat H^2 \rangle - \langle H \rangle ^2}.
\end{equation}

Also observe that under this non-linear extension, quantum states $|\Psi\rangle$ are said to be in equilibrium if $\left[ \hat H, \hat \rho \right]=0$ and $\mathbb{D} =0$. If $\left[ \hat H , \hat B \right] \neq 0$, their unitary evolution are limit cycles of the dynamics, and thus except for $\hat B = I$, all the equilibrium states and limit cycles are unstable according to Lyapunov \cite{Beretta:2009rmp}.\\

\subsection{Non-dissipative states and equilibrium}
Under this dynamics, all the solutions for $\mathbb{D} = 0$ are called non-dissipative, and it follows from Eq.~(\ref{eq:dissp}) that

\begin{equation} \label{eq:equilnrho}
    \hat{B} \ln \hat{\rho} = \beta \left( \langle H \rangle - \frac{\langle S \rangle}{\beta} \right) \hat{I} - \beta \hat{H},
\end{equation}
where $\hat{I}$ is the identity operator. By considering $\hat{B} \ln \hat{\rho}$, we effectively restrict $\ln \hat{\rho}$ to the support of $\hat{\rho}$, making the expression finite and well-defined. This is equivalent to replacing $\hat{B} \ln \hat{\rho} = \ln (\hat{\rho} + \hat{I} - \hat{B})$. In this way, the operator $\hat{B}$ eliminates contributions from the null space where $\ln \hat{\rho}$ is undefined. Let us define a modified operator:
\begin{equation}
\tilde{\rho} = \hat{\rho} + \hat{I} - \hat{B},
\end{equation}
which makes $\tilde{\rho}$ invertible and ensures that $\ln \tilde{\rho}$ is finite. Thus, Eq.~(\ref{eq:equilnrho}), can be written as
\begin{equation}
    \ln \tilde{\rho} = -\beta_{\text{eq}} \hat{H} - \ln Z,
\end{equation}
where $Z = e^{-\beta_{\text{eq}} \langle \hat f \rangle }=\text{tr} \left( \hat B\exp \left( -\beta_{\text{eq}} \hat{H} \right) \right)$ is the partition function. By exponentiation and multiplying on the left and right by the projector $\hat{B}$ to ensure that $\hat \rho$ is restricted to its rank, the density matrix at partial equilibrium takes the form\footnote{ Notice, that the Born rule, recovered in the Lindblad formulation at late times \cite{Weinberg:2016wh}, can  emerge under this dynamics through entropy maximization within the partially canonical states.}
\begin{equation}
  \hat{\rho} = \frac{\hat{B} \exp \left( - \beta_{\text{eq}} \hat{H} \right) \hat{B}}{Z}\,.
\end{equation}

 This implies that the dissipative dynamics leads to the canonical (or partial-canonical) distributions, i.e., $\hat{\rho} = \sum_n |\Psi_n\rangle\langle\Psi_n|$ is given in terms of equilibrium states denoted $|\Psi_n\rangle$. Therefore, we can identify the related parameter $\beta$ as the inverse of the thermodynamic temperature. Furthermore, for such a cases the expectation values $\langle \Delta \hat H \Delta \hat S \rangle$ and $\langle (\Delta \hat S)^2  \rangle$ are no longer independent, and thus 
\begin{equation}
    \beta_{\text{eq}} = \left( \frac{\langle (\Delta \hat S)^2  \rangle}{\langle (\Delta \hat H)^2 \rangle} \right)^{1/2}.
    \label{eq:equi}
 \end{equation}
It is important to remark that equilibrium states for which dissipation is null, could have some associated entropy, contrary to a pure state, for which the entropy is zero, as it follows from Eq. (\ref{eq:equi}) for $\hat{S}=0$. Moreover we can immediately identify a (particularly important) configuration leading to this situation, i.e., having a non-pure state in equilibrium. This happens for a mixed state formed by orthogonal states. 

\subsection{Pure state of gravitons}
Let us now consider a single coherent state of gravitons (over the Minkowski space) as our whole quantum system. At this point it is important to emphasize two relevant conditions in our work:
\begin{enumerate}
    \item According to the framework we are considering, it is very important to have a scenario in which the graviton coexists with other states which actually are orthogonal to it or to a coherent state formed by them. This is precisely what does happen in string theory. The graviton belongs to the massless symmetric excited state in the Neveu-Schwarz$-$ Neveu-Schwarz (NS-NS) sector along with the $B$-field and the dilaton. In type II theories the Ramond-Ramond (RR) sector also contains massless fields which by construction are orthogonal to those in the NS-NS sector. Therefore, string theory provides a natural scenario in which there are orthogonal quantum states to the graviton, which could lead (we shall see that it does) an instability of a system formed purely by a coherent state of gravitons. 
    \item  As mentioned earlier,  it was shown in \cite{Dvali:2017eba} that the dS space is well represented by a quantum coherent state of gravitons. This representation allows to compute the time-scale after which the quantum evolution departs from the classical mean field evolution described by the dynamics of the coherent state of gravitons. This scale is called the quantum time-break and it is inversely proportional to the number of particle species. The reason behind the fact that the quantum evolution can not be longer described by a classical theory after some time $t_Q$ is due to the scattering process among different quantum states and in consequence undergoes decoherence. Within the SQEAT framework, the quantum evolution is effectively described by the dissipation term $\hat{\mathbb{D}}$ in  Eq.~(\ref{eq:eom}). Therefore, we expect that the quantum process of scattering would increase the entropy driving the system into a mixed state.
\end{enumerate}

Based on the above expectations, it seems viable to effectively describe the quantum evolution of a coherent state of gravitons under the SEAQT framework and interpret the results in terms of the stability of a dS space. Then, let us consider a coherent states of gravitons, denoted $|\Phi_g\rangle$, given by
\begin{equation}
    | \Phi_g \rangle = \exp \left( \sum_{n=0} \epsilon_{\mu \nu} S^{\mu\nu}_{-n} \right) |0, k \rangle_{L}\otimes|0, \tilde{k}\rangle_R,
\end{equation}
where the symmetric operator is 
\begin{equation}
 S^{\mu \nu}_{-n} = \frac{1}{\sqrt{2}} \left( \psi_{-(n+\frac{1}{2})}^{\mu} \tilde \psi_{-(n+\frac{1}{2})}^{\nu} + \psi_{-(n+\frac{1}{2})}^{\nu} \tilde \psi_{-(n+\frac{1}{2})}^{\mu}\right),
\end{equation}
with $\psi^\mu_{-(n+\frac{1}{2})}$ and $\tilde{\psi}^\mu_{-(n+\frac{1}{2})}$ being the fermionic operators for the left and right modes respectively in the NS-NS sector.\\

We want to study the stability of this state in the presence of the orthogonal other states in string theory. For that we start by analyzing its stability and considering it as a single state in our system (i.e., for the moment we do not consider the presence of the other coherent states formed by the B-field and the dilaton).\\

Since our system is solely formed  by a coherent state of gravitons, let us consider the pure state $| \Phi_g \rangle \langle \Phi_g |$. Then the partition function is just
\begin{equation}
    Z = \langle \Phi_g| e^{-\beta_{\text{eq}} \hat H} | \Phi_g \rangle \,,  
\end{equation}
thus, if the operator $\hat{B}$ is given by $\hat B= | \Phi_g \rangle \langle \Phi_g | $,  the associate density matrix at partial equilibrium is 
\begin{equation}
    \hat \rho = \frac{| \Phi_g \rangle \langle \Phi_g | e^{-\beta_{\text{eq}} \hat H} | \Phi_g \rangle \langle \Phi_g |}{Z}
\end{equation}
which matches with the pure state density matrix. 
Therefore the matrix operator is actually a pure state of the form
$\hat \rho = | \Phi_g \rangle \langle \Phi_g |$.
Thus, under the non-linear dynamics, the pure state generated by the coherent states of gravitons, corresponds to partially canonical states. Such states, have zero entropy since represents states of complete knowledge, and are non-dissipative. Furthermore, since are proyectors of eigenvectors of the Hamilonian, leads to a zero contribution of the symplectic term in the SEAQT equation of motion, and in  consequense it dynamics is trivial, i.e., $\frac{d\hat \rho}{dt} = 0$. However, as we shall see, this solution is unstable in the Lyupanov sense, since any perturbation leads to a evolution of the density matrix. 

\section{Instability of a coherent state of gravitons in string theory}

In string theory, the graviton states are part of the NS-NS sector together with the B-field and the dilaton. As shown in the Appendix A, these states are orthogonal to each other, as well as any coherent states formed by them, i.e., coherent states of the B-field and the dilaton separately. This structure is actually the one that under the SEAQT framework allows to study the dynamical evolution of a system formed by a mixture state of coherent states. Therefore, we shall depart from a dS space constructed by a coherent state of gravitons which are slightly perturbed into a mixed state which takes into account the presence of the other coherent states in the NS-NS and RR sectors and analyze under which conditions this system reaches a thermodynamical equilibrium.\\

\subsection{Perturbations}
Consider an initial system formed by a coherent state of gravitons such that $\hat{\rho}=|\Phi_g\rangle\langle\Phi_g|$ as well as coherent states denoted $|\Phi_i\rangle$  constructed from the rest of fields in the NS-NS and RR sectors.  Notice that these states are orthogonal to the coherent state of gravitons (see Appendix A).  After some time we expect some interaction of these modes such that the system is perturbed  the presence of these modes. Therefore, after some perturbation of the pure state, the matrix density operator is given by
\begin{equation}
    \hat \rho = (1-\epsilon)| \Phi_g \rangle \langle \Phi_g | + \sum_{i}^{n} \delta_i | \Phi_i \rangle \langle \Phi_i |,
\end{equation}
 where $n$ refers to the number of coherent states $|\Phi_i\rangle$ orthogonal to the graviton and $\epsilon \ll 1$. To preserve the trace of the density matrix, we demand that $\sum_{i} \delta_i = \epsilon$ . In this sense, the dissipative term in the equation of motion Eq.~(\ref{eq:eom}) is non-zero. We proceed to compute all terms in  Eq. (\ref{eq:dissp}). Notice that the n-power of the perturbed density matrix is 
\begin{equation}
    \hat \rho^n = (1-\epsilon)^n |\Phi_g  \rangle \langle \Phi_g | +\sum_{i} \delta_i^n | \Phi_i \rangle \langle \Phi_i |.
\end{equation}
Thus, employing the replica trick we aim to approximate the term $\hat \rho \hat S$, by expanding the logarithmic terms for small $\epsilon$  giving us
\begin{equation} \label{eq:soper}
    \hat \rho \hat S \approx -\epsilon \hat B |\Phi_g  \rangle \langle \Phi_g |  - \sum_{i} \left( \delta_i \log \delta_i \right) \hat B | \Phi_i \rangle \langle \Phi_i |.
\end{equation}
Thus, the average expectation value for the entropy operator is
\begin{equation}
    \langle S \rangle = - \epsilon -\sum_i \delta_i \log \delta_i \,.
\end{equation}
Next, we construct the free energy operator $\hat f$, which combines the contribution of the Hamiltonian $\hat H$, and the entropy operator $\hat S$, as well as the thermodynamical parameter $\beta$, given in Eq. \ref{eq:freeop}. We find
\begin{equation} \label{eq:free_op}
    \hat f = \left( 1-\epsilon \right) \hat H | \Phi_g \rangle \langle \Phi_g | + \sum_i \delta_i \hat H |\Phi_i \rangle \langle \Phi_i |+ \frac{1}{\beta} \left( \epsilon \hat B | \Phi_g \rangle \langle \Phi_g | + \sum_i \delta_i \log \delta_i \hat B | \Phi_i \rangle \langle \Phi_i |\right),
\end{equation}
whereas its expectation value is given by
\begin{equation} \label{eq:free_av}
    \langle f \rangle = (1-\epsilon) \langle H \rangle_g + \sum_i \delta_i \langle H \rangle_i + \frac{1}{\beta} \left( \epsilon \langle \hat B \rangle_g + \sum_i \delta_i \ln \delta_i \langle \hat B \rangle_i\right),
\end{equation}
where we use the shorthand $\langle \cdot \rangle_g = \langle \Phi_g | \cdot | \Phi_g \rangle $ , where $\cdot$ is either the Hamiltonian operator, or the $\hat B$, operator  and $\langle \cdot \rangle_i = \langle \Phi_i | \cdot | \Phi_i \rangle $, implies the expectation value for the coherent states $| \Phi_i \rangle$.  The quantity $\beta$ is now a function on all $\delta_i$ and it is approximately (up to linear terms in $\epsilon$) given by 

\begin{equation}
    \beta[\delta]\approx\frac{-\sum_i\delta_i\log \delta_i\left(\langle H\rangle_i +\langle H\rangle_g\right)}{{\langle H^2 \rangle_g - \langle H \rangle^2_g -\epsilon \langle H^2 \rangle_g +2 \epsilon \langle H \rangle_g^2 +\sum_i \delta_i \langle H^2 \rangle_i-2 \langle H \rangle_g \sum_i \delta_i \langle H \rangle_i}}.
\end{equation}

For latter purposes, consider the contribution of the free energy operator in the graviton sector
\begin{equation}
    \langle \Phi_g | \hat f | \Phi_g \rangle = (1-\epsilon ) \langle H \rangle_g + \frac{1}{\beta[\delta]} \epsilon \langle B \rangle_g,
\end{equation}
as well as the contribution from the B-field and dilaton sectors,
\begin{equation}
      \langle \Phi_i | \hat f | \Phi_i \rangle = \delta_i\langle H \rangle_i + \frac{1}{\beta[\delta]} \delta_i \log \delta_i \langle B \rangle_i\,. 
\end{equation}

Using, the free energy contributions and the SEAQT framework, we derive the evolution equations for $\epsilon$ and $\delta_i$. This is done by direct substitution of the free energy operator, given by Eq. (\ref{eq:free_op}) and the average free energy, given by the Eq. (\ref{eq:free_av}). This is in general a complicated differential equation which contains the contribution of the perturbations $\epsilon$ and $\delta_i$. To separate each contribution, we can take the separate contribution of the average value of the free energy in the graviton sector as well in the orthogonal states, this allows us to write the evolution of the perturbations as
\begin{equation}\label{eq>epsilon}
    \frac{d\epsilon}{dt} = -\frac{\beta[\delta]}{\tau_D} \left( \langle \Phi_g | \hat f | \Phi_g \rangle -(1-\epsilon) \langle f \rangle  \right),
\end{equation}
as well as 
\begin{equation}
\label{eq:deltaeom}
  \frac{d \delta_i}{dt} =\frac{\beta[\delta]}{\tau_D} \left( \langle \Phi_i | \hat f | \Phi_i \rangle -\delta_i \langle f \rangle \right).
\end{equation}
To further simplify, we expand the evolutions equations to first order in the perturbations. After simplifying, this reduces to
\begin{align}
    \frac{d \epsilon}{dt} &= -\frac{1}{\tau_D} \bigg( \epsilon \beta[\delta]  \langle H \rangle_g  -\beta[\delta] \sum_i \delta_i \langle H \rangle_i -    \sum_i \delta_i \log \delta_i   \bigg).
\end{align}
This equation can be reduced to a dynamical evolution of perturbations $\delta_i$ in the form
    \begin{equation}\label{eq:equilibrium}
  \frac{d \delta_i}{dt} =-\frac{\delta_i }{\tau_D} \bigg(  \beta[\delta] \Delta E_i -\log \delta_i \bigg) \,,
\end{equation}
with $\Delta E_i= ( \langle H \rangle_g-\langle H \rangle_i)$. Although this is a difficult equation to analytically solve, we can infer some interesting features. For instance, if the  difference on energy $\Delta E_i$ is zero, meaning that the scattering between different coherent states are given among those with the same mean energy value, the system is unstable since the perturbation $\delta_i$ grows super-exponentially as
\begin{equation}
    \delta_i(t)\sim \exp\bigg(\exp(t/\tau_D)\bigg).
\end{equation}

However, we can see that in this case, the perturbations velocity growing depends solely on the parameter $\tau_D$. For large values of $\tau_D$ the perturbation grows very slowly. Actually, one can see that even if the scattering process is almost stopped, with a very small dissipation, the parameter $\tau_D$  determines how rapid the system becomes unstable. Since it seems that $\tau_D$ is a characteristic of the system not belonging to the quantum process, $\tau_D$ must be related to some classical behavior of the theory. In our case, and following \cite{Dvali:2017eba}, we identify $\tau_D$ with the {\it classical break time} of the dS space, given by $t_{cl}\sim 1/H$. This is indeed a strong assumption which requires a more profound analysis but provides an effective way to compare it with the time the system takes to undergo decoherence due to quantum processes. \\

\subsection{Reaching the equilibrium}
We have now all the necessary expressions to determine wether there are conditions upon which a coherent state of gravitons evolves reaches thermodynamical equilibrium as a mixed state. From Eq.~(\ref{eq>epsilon}) we see that if the perturbation $\epsilon$ grows over time, the final state would be a mixed state among the coherent states in the NS-NS and RR sectors. On the contrary, if $\epsilon$ decreases in size, we can say that the system reaches the equilibrium by restoring coherence on the graviton states.\\

From Eq.~(\ref{eq:equilibrium}) we see that the equilibrium would be reached when $d\delta_i/dt\sim 0$, this is when $|\beta[\delta_i]\Delta E_i-\log\delta_i|\ll 1$. Observe that such a condition is possible to reach, since $\beta[\delta_i]\sim \delta_i\log\delta_i$ for a finite (and small) number of orthogonal coherent states to the graviton. Therefore it is possible to get the necessary equilibrium condition such that, after some time $\beta[\delta_i]\Delta E_i\sim \log\delta_i$.\\

Let us then study the required conditions around this special region where equilibrium is close to be reached. Define  $F(\delta_i, \Delta E_i)=\beta[\delta]\Delta E_i-\log\delta_i$ and let us express it as 
\begin{equation}
    F(\delta_i, \Delta E_i)=k_i+ G(\delta_i, \Delta E_i),
\end{equation}
where $k_j$ is a very small constant indicating that the velocity of change in the perturbation is very close to zero. Therefore, near equilibrium, the function $G$ is considered to be null. However, far from equilibrium, $G(\delta_i, \Delta E_j)$ must contain the dynamics and particularities of how fast the perturbation evolves.\\

Then, in the general case, the solution of Eq.~(\ref{eq:deltaeom}), is given by
\begin{equation}
    \delta_i(t)\approx \delta_i(0) \exp\bigg(-\frac{k_i t}{\tau_D}\bigg)\exp\bigg(-I(\delta_i,\Delta E_i)\bigg),
\end{equation}
where $\delta_i(0)$ is the initial value of the perturbation and
\begin{equation}
    I(\delta_i,\Delta E_i)=\int \frac{G(\delta_i, \Delta E_i)}{\delta_i F(\delta_i,\Delta E_i)} d\delta_i.
\end{equation}

At the time when the system is almost in equilibrium, $G(\delta_i, \Delta E_i)\approx 0$ and we can say that
\begin{equation}
    \delta_i(t)\approx \delta_i(0) \exp(-k_i t/\tau_D).
\end{equation}

Thus
\begin{itemize}
    \item If \(k_i < 0\): The perturbation \(\delta_i(t)\) grows exponentially over time. This indicates that the system is unstable with respect to perturbations in the \(i\)-th state, and the density matrix evolves towards a mixed state, consisting in all other states in the NS-NS sector, as well as the RR one in the case of type II string theory.
    \item If \(k_i > 0\): The perturbation \(\delta_i(t)\) decays exponentially, and the system returns to the pure graviton state, indicating stability against perturbations in that direction.
\end{itemize}

Notice that the time required to reach the equilibrium, rather the system evolves to a mix state or goes back to a pure state of gravitons, is of the order
\begin{equation}
    t\sim \frac{\tau_D}{|k_i|}.
\end{equation}
Since the origin of such evolution is given by quantum scattering among the states of the string, we propose to identify this  time as {\it the quantum break time} $t_Q$. Together with the identification of the classical break time $\tau_D\sim 1/H$, we find that
\begin{equation}
    t_Q\sim \frac{1}{|k_i| H},
    \end{equation}
meaning that the factor connecting both time scales are given basically by $k_i$ which according to \cite{Dvali:2017eba} can be interpreted as a quantum coupling which powers suppress classical non-linearities corrections. For a small $k_i>0$, i.e. when the system is closed to equilibrium conformed by a mixture of coherent states, $t_Q>t_{cl}$. This seems to be in agreement with the Swampland conjectures about the impossibility of having a stable dS space at the quantum level. We go further in some details and possible different scenarios in our final comments.\\


\section{Final comments and conclusions}
The dynamics of a coherent state of string gravitons within the effective framework of the Steepest-Entropy-Ascent Quantum Thermodynamics (SEAQT) is described. Coherent states of gravitons, being pure states with zero entropy, are inherently unstable under the non-linear dynamics prescribed by the SEAQT equation of motion. This instability arises since any small perturbation induces an increase in entropy. This is described by a dissipative term in the SEAQT equation driving the system toward states of higher entropy and moving it away from the initial coherent state.\\

As the system evolves, it naturally progresses toward a stable equilibrium state that is a statistical mixture of  states available in the theory. For the case of (type II) superstring theories, the massless available states are contained in the so-called NS-NS and RR sectors.  Under the non-linear SEAQT dynamics, these fields would also contribute to the system's evolution toward a thermodynamic equilibrium state. This equilibrium would involve a mixture of all relevant massless modes, including both bosonic and fermionic degrees of freedom inherent in supersymmetric theories. The mixing of these states would lead to stable configurations that are consistent with supersymmetric backgrounds, and further supporting (or not) the swampland conjecture's implications in the context of superstring theory.\\


By identifying the coherent state of gravitons with the semi-classical limit of the dS space (in the weak gravity limit) as done in \cite{Dvali:2017eba}, we have a framework in which we can effectively study the stability of the dS space and infer the time scale it takes the system to undergo decoherence. Under the SEAQT framework, the coherent states evolve by increasing the associated entropy resembling the quantum scattering among gravitons and all other fields. In this sense, by considering a small dissipation, or equivalently, a small rate of scattering, a perturbation of the original coherent pure state grows over time, rendering the system unstable. The velocity of such evolution is encoded in the effective time $\tau_D$. Since its value seems to be independent on the quantum processes, we propose to identify it to the classical break-time associated to the dS space, this is $t_D\sim 1/H$, with $H$ the Hubble constant \cite{Dvali:2017eba}.\\

We also studied the conditions upon which a thermodynamical equilibrium is reached. This happens when $k_i=\beta[\delta]\Delta E_i- \log\delta_i\sim 0$, where $\delta_i$ is the perturbation along one of the orthogonal directions of the graviton (i.e., a perturbation driven by the dilaton and the B-field), and $\Delta E_i$ is the difference of the energy mean values related to the state of gravitons and the orthogonal ones and $\beta[\delta]$ is the effective inverse temperature associated during the dynamical process into equilibrium. Hence if the product of this effective temperature with the energy difference $\Delta E_i$ grows as $\log\delta_i$, the system would approach the equilibrium. As we mentioned, there are conditions to expect this would happen. Hence, near the equilibrium we can see that the system is unstable for $k_i<0$ and the time scale in which the system undergo to decoherence into a mixed state is given by $t\sim k_i/\tau_D\sim 1/(k_i H)$. We propose to identify this scale with the quantum break-time $t_Q$, since essentially depends on the quantum scattering between gravitons and orthogonal states producing a non-zero value for $\Delta E_i$.\\

Finally we observe that for divergent $\beta\Delta E_i$ (with a  almost zero thermodynamical temperature indicating the closeness to equilibrium) the coherent state of graviton is unstable, reaching the equilibrium in mixed states conformed of massless fields in the NS-NS and RR sectors of type II string theories. In that case
\begin{equation}
t_Q> t_{cl},\nonumber
\end{equation}
suggesting that quantum breaking should not occur since the theory, at the classical level, should provide with a mechanism which destroy the dS space before the quantum processes. This is in agreement with the dS-conjectures, which locates  the dS space as part of the so-called string Swampland.\\

The Swampland conjectures posit that certain low-energy effective field theories cannot be consistently embedded into a quantum theory of gravity, effectively "swamping" them out of the landscape of viable theories. Specifically, it suggests through the dS-conjectures, that stable dS (dS) vacua are difficult to achieve in string theory, while Minkowski and AdS vacua are more natural. The tendency of the SEAQT dynamics to favor equilibrium states corresponding to Minkowski or AdS backgrounds aligns with this aspect of the swampland conjecture, reinforcing the idea that only certain spacetime configurations are consistent with quantum gravity.\\

In conclusion, the SEAQT dynamics is compatible with expected in the swampland conjecture and provides valuable insights into the interplay between quantum thermodynamics and string theory. The instability of coherent states under the SEAQT framework and the resulting evolution toward an equilibrium mixture of massless modes are consistent with the principles of thermodynamics, particularly the tendency of systems to evolve toward states of maximum entropy. This behavior aligns with string theory phenomenology, where Minkowski and AdS spacetimes emerge naturally as stable backgrounds. Moreover, it offers support for the swampland conjecture by demonstrating how non-linear dynamics favor certain spacetime configurations over others.\\
\vspace{1cm}
\begin{center}
    {\bf Acknowledgments}
\end{center}
We thank Nana Cabo-Bizet for useful discussions around these topics. This work was supported by CONAHCYT throught the project CF-2023-I-682.

\appendix
\setcounter{section}{0} 

\section{Orthogonality among coherent spaces in the NS-NS sector}

For completeness, we show that a coherent state of fields formed by massless and massive states created by symmetric, antisymmetric and scalar operator in the NS-NS sector, are indeed orthogonal to each other. We elaborate on the case of a coherent state of gravitons and one constructed by states corresponding to the B-field. This means that together with the B-field, we also consider all massive modes created by the antisymmetric operator. Extension to the dilaton and states in the RR sector for type II superstrings are commented at the end.\\

Consider the algebra of the annihilation and creation operators for the NS-NS sector, namely
\begin{align} \label{eq:original}
    \left[ \psi_m^\mu, \psi_n^\nu \right] &= \left[ \tilde \psi_m^\mu, \tilde \psi_n^\nu \right] = m \delta_{m+n,0}\eta^{\mu \nu} \\ \nonumber
    \left[ \psi_m^\mu, \tilde \psi_n^\nu \right] &= 0
\end{align}
for $n\in \mathbb{Z}+\frac{1}{2}$. Then, the graviton coherent state is defined as
\begin{equation}
| \Phi_g \rangle = \exp \left( \hat C+\hat S \right) | 0 , k^\mu \rangle \,.
\end{equation}
where the operators are defined as
\begin{equation}
 \hat C = \sum_{n=1}^\infty \epsilon_{\mu \nu} \psi_n^{\mu} \tilde \psi_n^\nu  \,,\quad \text{and}\quad  \hat S = \sum_{n=1}^\infty \epsilon_{\mu \nu} \psi_{-n}^{\mu} \tilde \psi_{-n}^\nu.
\end{equation}
 $\hat{S}$ is an operator that creates symmetric states with respect to $\mu$ and $\nu$. Notice that the presence of both left-moving and right-moving operators is required due to the level-matching.\\

The commutation relations allow us to compute the action of the annihilation operator on the graviton coherent state $| \Phi_g \rangle$, which must be an eigenstate of the annihilation operator, this is
\begin{equation}
    \hat C | \Phi_g \rangle = \hat C \exp \left( \hat C+\hat S \right) | 0, k^{\mu} \rangle = \lambda | 0, k^{\mu} \rangle
\end{equation}
for some real number $\lambda$. This expression can be conveniently written as
\begin{equation}
\hat C | \Phi_g \rangle = - \left[ \hat C , \exp \left( \hat C+\hat S \right) \right] | 0, k^{\mu} \rangle
\end{equation}
since $\hat C$ annihilates the vacuum. Expanding this expression leads us to
\begin{align}
\hat C | \Psi_{\epsilon} \rangle &= C \exp (\hat C + \hat S) | 0, k^\mu \rangle \nonumber \\
&= \exp (\hat C + \hat S)  \left( \hat C + [\hat C, \hat C + \hat S] + \frac{1}{2!} [\hat C, [\hat C, \hat C + \hat S]] + \cdots \right) \lvert 0, k^\mu \rangle.
\label{eq:B_action}
\end{align}
Thus, the commutator between the annihilator operator and the sum of creation and annihilator operators appears in all terms. This commutator can be explicitly written in therms of the operators $\psi$ and $\tilde \psi$ as
\begin{align}
[\hat C, \hat C + \hat S] &= \sum_{n=1}^{\infty} \psi_n^\mu \tilde{\psi}_n^\nu \epsilon_{\mu \nu} \sum_{m=1}^{\infty} \left( \epsilon_{\rho \sigma} [\psi_n^\mu \tilde{\psi}_n^\nu, \psi_{-m}^\rho \tilde{\psi}_{-m}^\sigma] - \epsilon_{\rho \sigma} [\psi_n^\mu \tilde{\psi}_n^\nu, \psi_m^\rho \tilde{\psi}_m^\sigma] \right).
\end{align}

Using the commutation relations Eq.~(\eqref{eq:original}), only terms with $m = n$ survive, allowing us to write it as
\begin{align}
[\hat C, \hat C + \hat S] &= \sum_{n=1}^{\infty} \psi_n^\mu \tilde{\psi}_n^\nu \epsilon_{\mu \nu} \left( \epsilon_{\rho \sigma} n \eta^{\mu \rho} \tilde{\psi}_n^\nu \tilde{\psi}_{-n}^\sigma + \epsilon_{\rho \sigma} n \eta^{\nu \sigma} \psi_n^\mu \psi_{-n}^\rho \right).
\end{align}

However, since $\psi_n^\mu \lvert 0, k^\mu \rangle = \tilde{\psi}_n^\nu \lvert 0, k^\mu \rangle = 0$ for $ n > 0 $, higher-order commutators vanish when acting on the vacuum. Therefore, the series truncates, and we obtain
\begin{align}
\hat C \lvert \Psi_{\epsilon} \rangle = \sum_{n=1}^{\infty} n \epsilon_{\mu \nu}^{(n)*} \epsilon^{(n)}_{\mu \nu} | \Psi_{\epsilon} \rangle,
\end{align}
where the polarization tensor $\epsilon_{\mu \nu}$ must be symmetric, traceless, and satisfy transversality conditions $k^\mu \epsilon_{\mu \nu}^{(n)} = 0$. Furthermore, the coherent states are orthogonal to each other. This result follows from the symmetric (for the graviton) and antisymmetric (for the \( B \)-field) properties of the creation operators \( S^{\mu \nu}_n \) and \( A^{\mu \nu}_n \), respectively. Their commutator vanishes:
\begin{equation}
\left[ S^{\mu \nu}_n, A^{\rho \sigma}_m \right] = 0,
\end{equation}
since symmetric and antisymmetric operators commute due to their differing symmetry properties. This allows us to compute the inner product \( \langle \Phi_g | \Phi_B \rangle \) as
\begin{align}
\langle \Phi_g | \Phi_B \rangle &= \langle 0 | \left( \exp \left( \hat B + \hat S \right) \right)^\dagger \exp \left(\hat B + \hat A\right) | 0 \rangle =0,
\end{align}
showing that indeed, a coherent state of gravitons is orthogonal to one formed by B-field states. The argument extends straightforward to the dilaton. This ensures the existence of orthogonal states to the graviton coherent state. Moreover, in the case of type II string theory (A or B), the existence of the RR sector, orthogonal to the states in the NS-NS sector by construction, allows to have more orthogonal states.


\end{document}